# TEVATRON PHYSICS[*]

JOHN WOMERSLEY

*Fermi National Accelerator Laboratory*
*Batavia, IL 60510, USA*
*E-mail: womersley@fnal.gov*

These lectures form a personal, and not necessarily comprehensive, survey of physics at the Fermilab Tevatron proton-antiproton collider. They cover detectors, analysis issues, and physics prospects for the current Tevatron run.

## 1. Introduction

### 1.1. *Hadron Colliders*

The term "hadron colliders" is used to describe accelerators where a beam of protons is collided with a contra-rotating beam of protons or antiprotons. The next decade belongs to these machines: first the Fermilab Tevatron, and (starting around 2008) the LHC at CERN. Historically, there has been a complementarity of capability between hadron colliders and electron-positron colliders. Hadron machines emphasize maximum energy and hence maximum physics reach for new discoveries, while the electron-positron machines tend to have lower center-of-mass energy but cleaner events and better capabilities for precision measurements.

Hadron Colliders have the advantage that protons can be easily accelerated to very high energies and stored in circular rings. They have the disadvantage that antiprotons, if used, must be collected from the results of colliding a lower energy, high intensity beam on a target. This can of course be avoided by building a proton-proton collider, at the cost of needing a second ring of magnets. They also have the disadvantage that protons are made of quarks and gluons, so the whole of the beam energy is not concentrated in a single point-like collision. Quarks and gluons are also strongly interacting particles, so the collisions are messy. Despite these problems, hadron colliders are the best way to explore the highest mass scales for new physics.

---

[*] Lectures presented at the TASI 2002 Summer School, University of Colorado, Boulder, CO, June 2–28, 2002.

### *1.2. The Tevatron*

The overarching question for the world high energy physics program is "what sets the mass scale of the weak interaction to be ~100 GeV?" This question is addressed solely by colliders operating at the energy frontier. In the 1990's, there were four such machines: the Tevatron collider (Run I), LEP at CERN, the SLC at SLAC, and HERA. In contrast, between 2002 and roughly 2008, the Tevatron is the only machine that can address the central problems of the field: LEP and SLC have closed, and HERA's reach is limited. Increased energy and a greatly increased luminosity at the Tevatron make possible a new round of experimentation.

The Tevatron physics program involves:
- Precise measurements of the known quanta of the standard model, to search for indirect hints (or constraints) on new particles or forces;
- Direct searches for new physics i.e. beyond the known standard model particles and forces.

The Tevatron program has the potential for a discovery that would change the direction of particle physics.

Between 1992 and 1995 (Run I), the Tevatron delivered about 120 pb$^{-1}$ to each experiment at a center-of-mass energy of 1.8 TeV. Run II started in 2001, with an increased center-of-mass energy of 1.96 TeV, using the new Fermilab Main Injector as part of the accelerator chain. We expect 2–4 fb$^{-1}$ in the first phase of Run II, after which detector upgrades will be needed; the goal is to have delivered 10–15 fb$^{-1}$ before the LHC becomes competitive.

### 2. Detectors

A typical detector surrounds the interaction point with concentric layers of particle measurement devices, as shown in Fig. 1.

The innermost parts of the detector measure charged particle trajectories in a magnetic field. Particle direction and momentum can be inferred. The innermost layers need to make the most precise measurements and use silicon wafers; these tend to be expensive, and so the outer layers use less precise but less costly technologies. Surrounding the tracker is a calorimeter. The calorimeter induces particle showers in dense material and measures the particle energies. It also distinguishes electrons and photons from hadrons by the shower topology. Finally, the outermost layers of the detector measure and identify muons. Combining all these measurements, it is also possible to infer the presence of non-interacting particles like neutrinos, since they leave unbalanced momentum in the transverse plane ("missing transverse energy", $E_T^{miss}$).

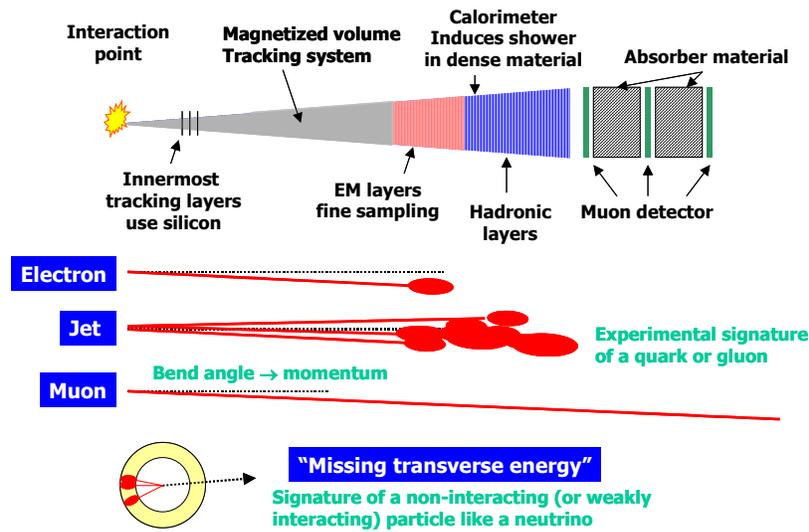

**Figure 1**: schematic of an azimuthal section of a tyrpical collider detector, and the particle signatures seen in it.

There are two large multi-purpose detectors at the Tevatron: CDF and DØ (D Zero). The detectors are complementary in that CDF detector places more emphasis on tracking measurements, while DØ emphasizes calorimetry. The physics reach of the detectors is not dramatically different: it tends to either be driven by cross sections (as in searches for SUSY) or the respective strengths of the detectors tends to balance (as for the top quark in Run I).

## 2.1. *Tracking*

The ability to identify b-quarks is very important at hadron colliders, as a signal for top, supersymmetry and Higgs. Experimentally, we can exploit the fact that a b-quark forms a B-meson which travels ~ 1mm from the collision point before decaying. To reconstruct this decay, we need to reconstruct tracks with a precision at the 10 μm level. We can tag the B-decay either by asking for two or three tracks having a significant "impact parameter" (distance of closest approach to the fitted primary vertex), or by explicitly reconstructing a secondary decay vertex separated from the primary. Figure 2 shows these two methods schematically. Both CDF and DØ use silicon detectors close to the beam pipe to achieve the necessary tracking precision. Silicon detectors use wafers of silicon of order 2 cm × 12 cm with ~50μm pitch between readout strips and on-board amplifier-discriminator chips. These "ladders" are arrayed in cylindrical layers around the beam pipe and supported on structures that provide rigidity and cooling. Both the CDF and DØ silicon detectors were assembled at Fermilab's Silicon Detector Facility. They are commissioned and working well in the collider. In CDF, the silicon detector feeds an impact parameter trigger which allows B-decay candidates to be selected online: a first

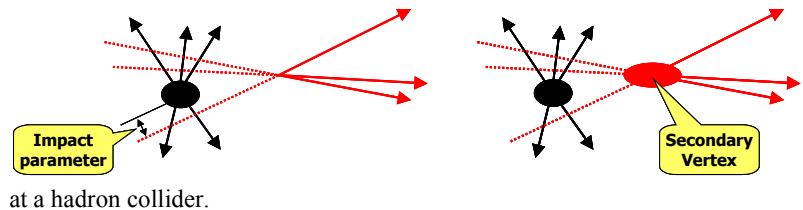

at a hadron collider.

**Figure 2**: Impact parameter and secondary vertex techniques for b-tagging.

The detectors use separate outer tracker systems to cover the tracking region beyond the silicon. The CDF Central Outer Tracker is a large drift chamber with 96 wire planes and 30,000 sense wires. The DØ Central Fiber Tracker uses 8 layers of 1mm diameter scintillating fibers, read out with cryogenic photodetector devices. Both the COT and CFT have the capability to provide triggers on high-transverse-momentum charged tracks.

### 2.2. *Calorimetry*

The basic tool for detection of electrons, photons and hadronic jets (the signatures of quarks and gluons) is a segmented calorimeter surrounding the tracking system. The basic idea is to induce a shower of interactions between the incident particles and a dense material, and to measure the ionization energy deposited in the shower. From this, the particle energy is deduced. For electromagnetically interacting particles ($e^{\pm}$, $\gamma$) the situation is straightforward. Above ~100 MeV, pair production and bremsstrahlung dominate the energy loss mechanisms. Shower development involves the processes $\gamma \rightarrow e^+e^-$ and $e^{\pm} \rightarrow e^{\pm}\gamma$ and it scales with the radiation length $X_0$. It is easy to show that the total charged track length in the shower, and hence the sum of ionization, is proportional to the incident energy E, and that the RMS is proportional to $\sqrt{E}$. Strongly interacting particles (hadrons) also cascade in matter, but many more processes are involved: as well as inelastic hadronic collisions, energy is lost to ionization, electromagnetic cascades from produced $\pi^0$ mesons that decay to photons, nuclear binding energy, neutrino production, and nuclear excitation. Hadronic showers scale with the nuclear interaction length $\lambda$: the showers are longer, wider, start later and have more fluctuations than an EM shower of the same energy. Because not all of the above processes are detectable, the response to a hadron is usually lower than to an electron of the same energy, the so-called $e/\pi$ ratio.

For reasons of cost and compactness, collider calorimeters are "sampling calorimeters" that alternate dense but inert absorber with layers of sensitive medium. A fixed fraction of the shower energy is sampled in the sensitive medium. In the case of CDF, the sampler is scintillator tiles. New forward calorimeters were added for Run II and these exploit wavelength-shifting fibers to give a compact and flexible way to route the light signals to photomultiplier tubes. DØ uses liquid argon as the sampling medium: the calorimeter consists of an absorber structure with electrodes, all of which is immersed in a vessel containing liquid argon, which serves as the ionization collector.

Calorimeter energy resolution is usually dominated by statistical fluctuations in the number of shower particles, which scale as $\sqrt{E}$. The fractional resolution is therefore often quoted as "x%/$\sqrt{E}$" (where E is assumed to be measured in GeV): typical real-world values are 15%/$\sqrt{E}$ for electrons, 50%/$\sqrt{E}$ for single hadrons and 80%/$\sqrt{E}$ for jets. Other terms contribute in quadrature: a "noise term", independent of E, which dominates at low E, and a "constant term", a constant fraction of E, which dominates at high energies and can come from calibration uncertainties, nonlinear response, and unequal response to hadrons and electrons.

At high energies, jets become very clear in hadron-hadron collisions, and the calorimeter gives a very visual picture of the event topology. It also allows electrons to be clearly distinguished from jets.

### 2.3. *Muon system*

The outermost part of the detectors is the muon system. Since all other types of charged particles are absorbed, muons can be identified as charged tracks that traverse absorber outside the calorimeter. The outside of the detector is therefore covered with large-area detector planes to identify and trigger on muons. In the case of DØ, an approximate stand-alone muon momentum measurement can be made using magnetized iron, but in both DØ and CDF the central tracker provides a precise measurement once the muon track has been linked to matching trace in the central tracker.

### 2.4. *Trigger*

The accelerator luminosity is set by the physics goals. For example, to find the Higgs we need $\sim 10 \text{fb}^{-1}$ of data; to do this in a reasonable time requires a collision rate of $2 \times 10^{32}$ cm$^{-2}$ s$^{-1}$. The total inelastic proton-antiproton cross section is huge compared to the processes we are interested in studying. Even with bunch crossings every 396 ns at the Tevatron, there is more than one inleastic collision each time the bunches cross. The triggering challenge is then the real-time selection of perhaps 50 events per second (the maximum we can store to tape) out of this collision rate of 2.5 million per second. We do this by rapid identification of high energy particles, and of comparatively rare objects (electrons, muons…). Triggering is perhaps the greatest challenge to experiments at hadron colliders.

Both CDF and DØ use a three-level trigger system to reduce the rate progressively. The first level selects 10–40 kHz of collisions based on fast information from specialized detectors. The second level reduces this to a few hundred Hz using microprocessors that perform fast calculations on a subset of the data and reject most of the level 1 candidate events. The third level uses a farm of computers, with access to the full event readout; again most of the candidates from the previous level are rejected and 50 Hz is stored to tape. These events are passed to a computer farm where the reconstruction program is run on them within a few days of data taking, followed by storage in a tape robot system.

The experiments operate with a "trigger list" of order 100 different trigger requirements at each trigger level, optimized for particular physics processes or final states. For many relatively high-rate processes, like jet production, we cannot run the trigger without restriction, because the cross section at low energy would exceed our capabilities. Such triggers are "prescaled", and only a known, fixed fraction of the events satisfying the trigger are saved.

**2.5.** *Simulation Tools*

A "Monte Carlo" is a Fortran or C++ program that generates events. Events vary from one to the next (random numbers are used), so we can expect to reproduce both the average behavior and the fluctuations of real data. Event generators may be parton level, which include just the parton distribution functions and the hard matrix element, or they may additionally include initial state radiation, final state radiation, hadronization of quarks and gluons, decays of unstable particles, and generation of the underlying event. Separate programs are used for simulation of the detector response to events. These are written by the detector collaborations but based on standard toolkits (GEANT being the best known and most widely used).

**2.6.** *Computing*

Computing for data processing and analysis is a challenge for modern experiments both because of the quantity of data and because of the size and distributed geography of the collaborations. In the past, this led to the invention of the World-Wide Web as a way to share information. Now the challenge is to share data and computing power. There is a natural synergy between the needs of our experiments and current ideas about "Grid" computing. The Tevatron experiments are already making something like a Grid a reality and are distributing their data for analysis using a Fermilab-developed system called SAM. They are also exploring ways for remote collaborators to assist in monitoring detector operations using web tools over the internet.

**3. Hadron-hadron collisions**

Hadron-hadron collisions are complicated by the fact that a hadron collider is really a broad-band quark and gluon collider. The incident particles to the parton-level hard scattering are selected from parton distributions at some factorization scale. The hard scattering is calculated at a given order in $\alpha_S$, and at some renormalization scale. There may be initial state radiation before the hard scattering, or final state radiation after it. Outgoing quarks and gluons fragment into jets of hadrons, and the remaining quark/gluon content of the interacting beam particles forms a soft "underlying event."

Since the incoming parton momentum fractions $x_1$ and $x_2$ are a priori unknown, and usually beam particle fragments escape down the beampipe, the longitudinal motion of the parton-parton center of mass cannot be reconstructed. We therefore focus on transverse variables. We use the transverse energy and momentum, $E_T = E \sin\theta$ and $p_T = p \sin\theta$, where $\theta$ is the angle between a particle direction and the beam. We also use longitudinally boost-invariant quantities, the primary one being the pseudorapidity $\eta = -\ln \tan(\theta/2)$. This maps on to angles in the detector: $\eta = 0, 1, 2$ being $\theta = 90°$, and roughly $40°$ and $15°$ respectively. For massless particles, pseudorapidity and true rapidity coincide. Particle production usually scales per unit of (pseudo)rapidity.

**4. Quantum Chromodynamics**

No one doubts that Quantum Chromodynamics (QCD) describes the strong interaction between quarks and gluons. Its effects are all around us: it is the origin of the masses of hadrons, and thus of the mass of stars and planets. This doesn't mean it is an easy theory to work with. As well as using hadron colliders to test QCD itself, we find that it is so central to the calculation of new physics processes, and their backgrounds, that we need to make sure we can have confidence in our ability to make predictions in this framework.

QCD is a gauge theory describing fermions (quarks) which carry an SU(3) color charge and interact through the exchange of vector bosons (gluons). It has the interesting features that the gluons are themselves colored, and that (as a consequence) the coupling constant runs rapidly and becomes weak at momentum transfers greater than a few $(GeV)^2$. These features lead to a picture where the quarks and gluons are bound inside hadrons if left to themselves, but behave like free particles if probed at high momentum transfer. This is exactly what was seen in the deep inelastic scattering experiments at SLAC in the late 1960's which led to the genesis of QCD. The short-timescale, hard scattering involves the electron scattering elastically off a single, pointlike constituent inside the nucleon; the other quarks do not participate. Afterwards (over a longer timescale) the scattered quark, knocked out of the proton, radiates lots of gluons and quark-antiquark pairs which combine with each other and the colored remnant of the proton to form colorless hadrons. This is called "fragmentation" or "hadronization." At high momentum transfers these hadrons form a jet of roughly collinear particles whose energy and direction correspond to that of the quark or gluon. This picture becomes very intuitive when compared with high-energy events from LEP. One can see the topologies of two jets corresponding to $e^+e^- \to \bar{q}q$ and three jets from $e^+e^- \to \bar{q}qg$. High energy events at the Tevatron are also simple to recognize (see Fig. 3).

Event generators model the hadronization process by dividing in into two phases: a perturbative phase, wherein a shower of quarks and radiated gluons develops and $Q^2$ evolves downwards; and a nonperturbative phase, once $Q^2 \sim 1$ GeV$^2$ and the coupling constant becomes strong, where those quarks and gluons are grouped into hadrons (which may then decay). The PYTHIA and HERWIG programs contain two popular Monte Carlo treatments of this process.

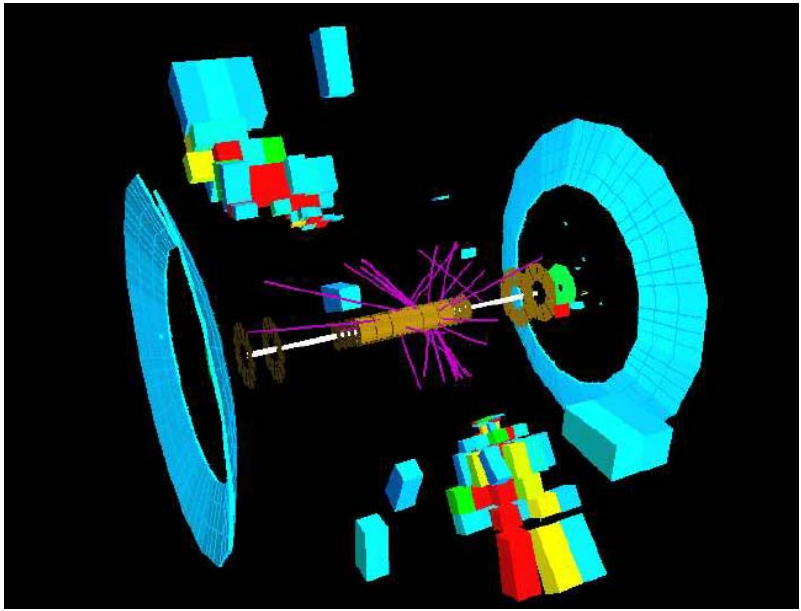

**Figure 3**: High energy dijet event in Run II, as recorded by DØ.

### 4.1. *Jets*

Experimentally, we cannot make the same identification between hadrons and their "parent" parton. We need a jet algorithm to identify jets and define their properties rigorously. We would like an algorithm that can be run both on partons and on final state particles, and on detector measurements, and yield comparable results in all cases: in other words, one that does a good job of integrating over the non-perturbative and hard-to-calculate hadronization process. It is important to define jets properly even at the parton level: a naïve one parton = one jet identification makes no sense for soft or collinear gluon radiation, for example.

The traditional choice at hadron colliders has been the "cone algorithm." Jets are defined as the sum of energy within a cone of radius $R^2 = \Delta\eta^2 + \Delta\phi^2$. The traditional choice at electron-positron colliders has been the successive recombination or $k_\perp$ algorithm; jets are formed by combining particles with low relative transverse momentum until only clusters well separated in transverse momentum remain. The transverse momentum measure used is $d_{ij} = \min(E_{Ti}^2, E_{Tj}^2)\,\Delta R_{ij}/D$. Typical jet sizes used are $R = 0.7$ in the cone and $D = 1$ in $k_\perp$.

One of the problems with the cone algorithm is that cone jets can overlap. In that case, how do we assign the shared energy? Are there two jets or one? We need a set of "split/merge" rules, which are unavoidably arbitrary. For example, in DØ, jets are merged if they share more than half the softer jet's energy and otherwise the overlapping pair is split into two jets. Merging and splitting turns out to be rare for high energy jets (above about 50 GeV at the Tevatron) but affects 15–20% of 20 GeV jets. A similar merging and splitting procedure then has to be applied to partons, which leads to the adoption of an ad hoc parameter called $R_{sep}$: two partons can form a jet if their separation is less than $R_{sep}$, otherwise they form two jets. $R_{sep}$ is chosen to be 1.3 times the jet cone radius, based on seeing how jets merge and split in real data when the merge/split rules are applied. The need for these ad hoc rules with cone jets is one motivation to use the $k_\perp$ algorithm. It has also been suggested that $k_\perp$ jets would have better energy resolution, but that is not yet clearly demonstrated in real life.

Jet calibration requires three steps. First, calorimeter stability and uniformity must be established. This requires pulsers or light sources to test the readout, and use of collider data (requiring azimuthal uniformity, or using muons as a source of known energy deposits). Secondly, the overall energy scale must be established. One can use testbeam data, or set the measured energy equal to the momentum for isolated tracks seen in the central tracker, or use resonances decaying into photons or electrons ($\pi^0$, J/$\psi$, Y, Z). One then adjusts the calibration to obtain the known mass. Finally, we must relate the energy scale for jets to that for electrons or photons. This can be done using Monte Carlo simulation of jet fragmentation (as in CDF) or by using photon+jet events and requring $E_T$ balance (done in DØ). It is customary to correct back to the "particle level" (in other words, the effects of the detector are removed but not those of hadronization) and also to remove energy that did not originate in the hard scattering (which means we need to subtract an estimate of the underlying event energy).

The jet energy resolution must also be established so that it can be deconvoluted from the measured distribution. This is done from collider data by looking at $E_T$ balance in two jet events. For two jets, we can define an asymmetry $A = (E_{T1} - E_{T2})/(E_{T1} + E_{T2})$; then the $E_T$ resolution $\sigma_{ET}/E_T = \sqrt{2}\,\sigma_A$. Since real events have additional soft jets and are not perfectly balanced in $E_T$, it is necessary to progressively tighten cuts and extrapolate to the limit of no soft radiation. Typical jet resolutions are in the range $75-100\%/\sqrt{E(\text{GeV})}$.

### 4.2. *Jet Cross Sections at the Tevatron*

In Run I, both CDF and DØ measured jet cross sections at 1.8 TeV using the cone algorithm ($R = 0.7$). The cross section falls by seven orders of magnitude from 50 to 400 GeV. Agreement with NLO QCD is generally good, except perhaps at the highest energy end of the spectrum. When these results first appeared, there was some excitement that the CDF data, especially, were significantly in excess of QCD for $E_T > 250$ GeV. This sparked a renewed investigation of uncertainies in the prediction arising from parton distributions, and it was found that the gluon distribution is not in fact well determined in this kinematic regime; a small enhancement in the gluon content is allowed and can model the CDF data very well. The latest PDF fits from CTEQ and MRST now make use of the Run I jet cross sections from DØ to pin down the gluon distribution at high x.

Both CDF and DØ have now started measuring jet energy distributions from Run II. CDF are making use of their new forward calorimetry to cover the whole range of pseudorapidity (see Fig. 4). Jet calibrations are not yet final, but already we see events with transverse energies beyond 400 GeV. With the full Run II dataset this will reach as far as 600 GeV, allowing us to pin down the high-energy behavior of the cross section and thus better determine the gluon content of the proton.

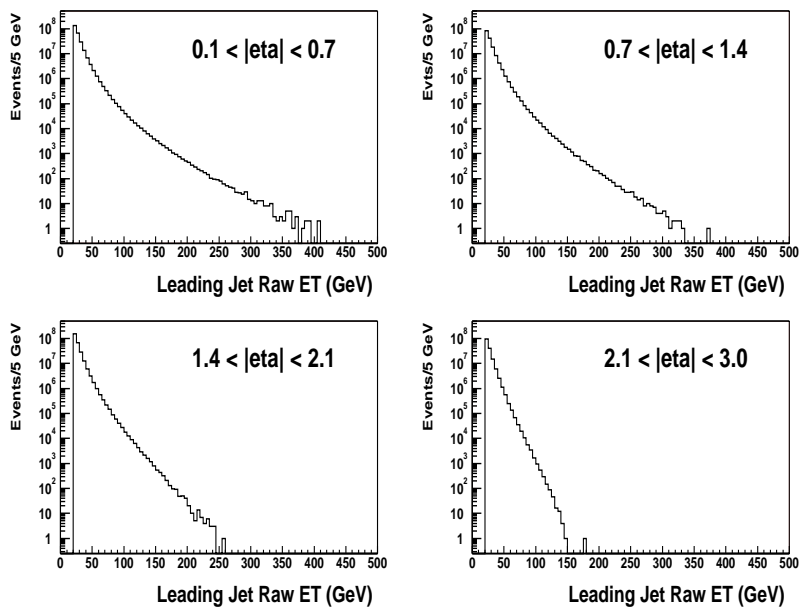

**Figure 4**: Preliminary Run II jet distributions from CDF.

DØ also measured the jet cross section in Run I data using the $k_\perp$ algorithm. The two algorithms yield different cross sections for collider data. Qualitatively this is what is expected, since the $k_\perp$ algorithm tends to gather more energy into jets. Quantitatively, however, it is not yet clear whether the differences are really as expected. We will try to address this question with early Run II data.

### 4.3. *Heavy Flavor Production*

Run I left many unanswered questions about heavy flavour (charm and bottom) production. Resolving these is important because many new particles result in heavy flavour signatures. The inclusive b-jet and B-meson production cross sections lie significantly above the QCD prediction (see Fig. 5), though it can be made to fit better using resummation and retuned fragmentation functions (from

LEP data). For charmonium, the measured cross section requires a large colour-octet component but that is not consistent with the observed J/ψ polarization. The CDF secondary vertex trigger in Run II is working beautifully, and the resulting huge charm and bottom samples will allow these puzzles to be explored in much more detail. DØ now has preliminary Run II J/ψ and muon+jet cross sections which are the first steps in measuring the charmonium polarization (and thus production process) and the b-jet cross section.

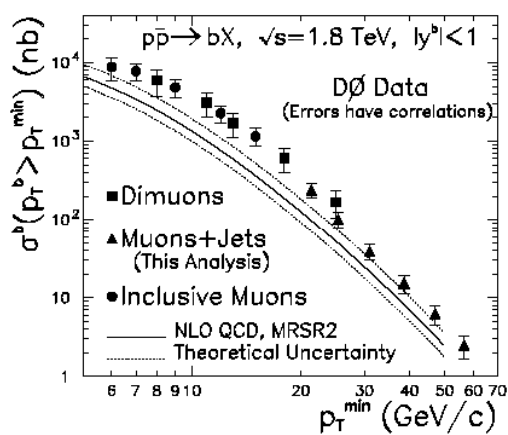

**Figure 5:** Inclusive b-jet cross section (DØ, Run I data) compared to the QCD prediction.

*4.4. Hard Diffraction*

Another QCD-related puzzle is hard diffraction. In this kind of event, a high-momentum-transfer collision occurs but one of the incoming beam particles appears to leave the collision intact, instead of being destroyed in the process. This observation is rather surprising and needs to be pinned down better, and related quantitatively with similar phenomena observed at HERA. Both CDF and DØ have new instrumentation for diffractive physics in Run II. This will allow us to test some of the basic assumptions on how to tag hard diffraction that have been used in earlier studies and will provide a sanity check for ideas of Higgs production through this mechanism at the LHC.

## 5. B-physics

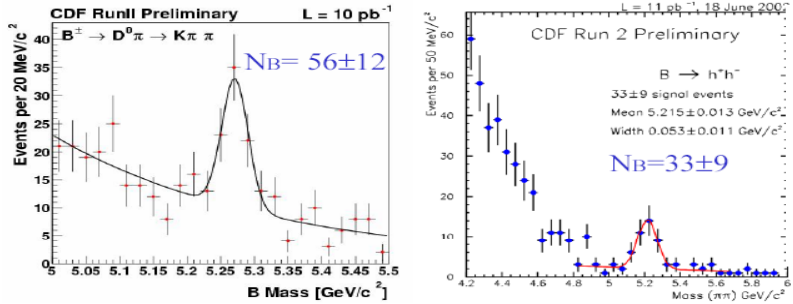

**Figure 6:** hadronic B-meson samples recorded by CDF in Run II using the SVT trigger.

The mixing between the three generations of quarks results in subtle violations of the CP symmetry relating particles and antiparticles. Understanding this symmetry will help explain why the universe is filled with matter, not antimatter. In the decays of B-mesons, these symmetry violations can be large, and so B-hadrons have become an important laboratory to explore the "unitarity triangle," which relates the elements of the Cabibbo-Kobayashi-Maskawa (CKM) quark mixing matrix. In Run II we want to confront the CKM matrix in ways that are complementary to the electron-positron B-factories. CP violation is now established in the B system through the decay $B_d \to J/\psi\, K_S$. The measured mixing angle is consistent with the standard model but, by itself, cannot exclude new physics. The BaBar and BELLE experiments can and will do much more with their data, but the Tevatron can uniquely access the $B_s$ meson, which is not produced at the B-factories and has therefore been called the "el Dorado" for hadron collider B-physics. By measuring the mixing rate between $B_s$ and $\bar{B}_s$, we can determine the length of one of the sides of the unitarity-triangle and complement the B-factories' measurements of its angles. CDF expect to be sensitive to Standard Model mixing with a few hundred inverse picobarns. It will also be interesting to see if there is sizeable CP violation in $B_s \to J/\psi\, \phi$ (it is expected to be small); while the decay $B_s \to KK$ at the Tevatron complements $B_d \to \pi\pi$ that is measured at the B-factories. Together they can pin down the unitarity triangle angle $\gamma$. There are many other opportunities, such as $\Lambda_b$ properties and searches for rare decays. CDF already have most impressive results from Run II, building on their Run I experience together with new detector capabilities (silicon vertex trigger and time of flight detector). Lepton-triggered signals for $B^{\pm} \to J/\psi\, K^{\pm}$, $B^0 \to J/\psi\, K^{*0}$, and $B_s \to J/\psi\, \phi$ are seen; while using the Silicon Vertex Trigger, the purely hadronic modes $B^{\pm} \to D^0 \pi \to K\pi\pi$ and $B \to$ hadron hadron are being recorded (see Fig 6). We can also look forward to CDF exploiting an enormous sample of charm mesons. In DØ the tools are being put in place for a B-physics program. The inclusive B lifetime has been measured and B mesons are being reconstructed. DØ does not exploit purely hadronic triggers but benefits from its large muon acceptance, forward tracking coverage, and ability to exploit $J/\psi \to e^+e^-$.

## 6. W and Z bosons

In Run II, we will complement our direct searches for new phenomena with indirect probes. New particles and forces can be seen indirectly through their effects on electroweak observables. The tightest constraints will come from improved determination of the masses of the W and the top quark.

Until the LHC starts operations, the Tevatron is the world's only source of W and Z bosons. Production is dominated by quark-antiquark annihilation. The typical boson $p_T$ is a few GeV, but about 10% of the W/Z bosons are produced with one or more jets of significant $E_T$ (>25 GeV). Though the dominant decays of W's and Z's are into jets, they are swamped by the enormous QCD jet cross section. This means that experimentally we focus on the leptonic decay modes of the W and Z.

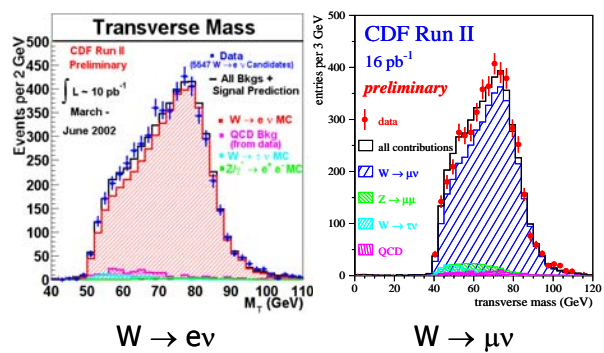

**Figure 7:** W transverse mass distribution in electron and muon channels (CDF, Run II)

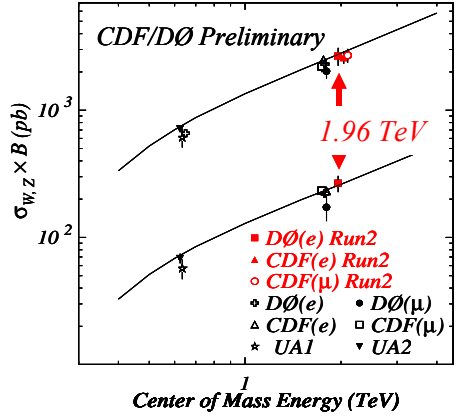

**Figure 8:** W and Z cross section measurements at proton-antiproton colliders.

W bosons are identified in the detector as a high-$p_T$ lepton together with a neutrino, identified as unbalanced transverse energy. We cannot reconstruct the longitudinal momentum of the neutrino but we can construct a transverse mass $m_T = \{2p_T^{lepton} E_T^{miss} (1-\cos\Delta\phi)\}^{1/2}$; the maximum transverse mass is equal to the the W mass. Transverse mass distributions from Run II are shown in Fig. 7. The W and Z cross sections can be measured extremely well and are in excellent agreement with the QCD prediction (Fig. 8). In fact it is likely we shall start to use the W cross section as an absolute luminosity normalization at some point in Run II, as it is easier both to measure and to predict than the total inelastic proton-antiproton cross section which is used at present.

One of the major goals of the Tevatron program is the W mass measurement. The simplest method is to fit the transverse mass distribution; recently we have started to fit the lepton $p_T$ and the $E_T^{miss}$ distributions as well, and combine the three results. The whole key is to understand the systematics: one must keep beating them down as the data demand more precision. One can use the Z to constrain many effects, such as the lepton energy scale, the boson production process and boson $p_T$. Currently, the W mass is known to be $m_W = 80\,451 \pm 33$ MeV; the measurement is dominated by LEP data. The Tevatron Run I results fixed the W mass at the 60 MeV level, but it will take a Run II dataset of order 1 fb$^{-1}$ before we can significantly improve the world knowledge of $m_W$ — not a short-term prospect. Given 2 fb$^{-1}$ we will be able to drive the uncertainty down to the 25 MeV level per experiment, with an ultimate capability of 15 MeV per experiment.

Both experiments now have preliminary results from their Run II samples of W and Z candidates. They have measured the cross sections at the Tevatron's new centre of mass energy of 1.96 TeV (Fig. 8) and used the ratio of the W to the Z to indirectly extract the W width. CDF have also taken a first look at the forward-backward asymmetry in $e^+e^-$ production in Run II.

### *6.1. QCD Aspects of W and Z production*

To describe the production of bosons with large $p_T$ (> 30 GeV), we can use perturbative QCD; NLO calculations exist. For small $p_T$ (< 10 GeV), the fixed order calculation fails, and one needs to resum large logarithms of $m_W^2/p_T^2$. At intermediate $p_T$, the two regions must be matched. This approach works well, but one needs to use data to extract a couple of non-perturbative parameters for the resummation calculation. We can use Z data to predict the W $p_T$ in this way.

W and Z production with jets is an important background to the top quark, to the Higgs, and to many new physics searches. CDF have measured the W+jet cross section as a function of jet $E_T$ and find good agreement with NLO QCD. For 3, 4 or more jets, only leading order calculations exist, and the CDF data agree reasonably well as long as the renormalization scale is chosen appropriately. DØ has used their W+jet sample to study color coherence. By comparing the pattern of energy flow around the (colorless) W to that around the jet, evidence for the QCD-predicted color interference effects in soft gluon emission is seen in the data.

Because leptons can be measured well, and the production process is well-understood, Drell-Yan lepton pair production at dilepton masses above the Z pole provides a sensitive test for new physics. Examples are searches for compositeness and for indirect effects of extra dimensions.

### 7. The Top Quark

The top quark is unique among the elementary constituents of matter because of its high mass (175 times the proton mass, compared to 5 times the proton mass for the next heaviest quark). In the standard model this means that it alone has a strong (non-perturbative) coupling to the Higgs boson. Is nature giving us a hint here? Whether the Higgs or something else turns out to be the origin of the electroweak symmetry breaking, the top quark seems to be uniquely connected to the mechanism of mass generation, and the Tevatron collider is the world's only source of top quarks.

The search for the top was pursued directly at $e^+e^-$ and $p\bar{p}$ colliders for many years, with mass limits increasing from $m_t$ >23 GeV in 1984 to $m_t$ > 131 GeV by 1994. In parallel, the increasingly precise electroweak parameter measurements made at LEP and SLC permitted indirect estimates of the top mass to be made, which by the early 1990's favored the mass range around 150–200 GeV. Direct and indirect measurements converged in 1995 when the top quark was discovered by CDF and DØ with a mass around 175 GeV.

Top quarks are produced in $t\bar{t}$ pairs at the Tevatron (electoweak single-top or tb production has not yet been observed, and will be discussed later). In the Standard Model, the top decays almost 100% to a W and a b quark. The most accessible signatures are those where one or both W's decays to an electron or muon, i.e. lepton + 4 jets + $E_T^{miss}$, or 2 leptons + 2 jets + $E_T^{miss}$. The all-hadronic (6 jets) channel has also been investigated but is challenging because of the large QCD backgrounds. Tagging the jets from the b-quarks, either using soft lepton tags or the silicon vertex detector, helps to pull the top signal out of the W+jets background.

Both DØ and CDF are on the road to "rediscovering" top for the spring 2003 conferences, and both experiments have candidate events. We can look forward to significant improvements in the short to medium term because the Run I dataset was so statistically limited—of order 20 clean events per detector. In Run II, we expect roughly 500 such clean events for every inverse femtobarn recorded.

### 7.1. *Top mass*

The top mass is a fundamental parameter of the Standard Model and critical input to electroweak fits. It affects the Standard Model prediction of the W and Higgs mass through radiative corrections.

In the lepton + jets channel, there is one unknown in reconstructing the event kinematics (the longitudinal momentum of the neutrino). There are three constraints (the lepton+neutrino and two of the jets should reconstruct to the W mass, and the top mass should be equal to the antitop). This allows a two-constraint kinematic fit, but there are up to 24 combinatoric ambiguities. The latter can be greatly reduced if one, or even better, two jets are b-tagged. In addition to the combinatorics, gluon radiation can add extra jets which worsens the mass resolution significantly. The basic fitting procedure is to evaluate a function *f* which is an estimator of the top mass (such as the result of a kinematic fit) for each event in the sample. The distribution of *f* is then compared to what is expected, given a hypothesised $m_t$, for a sample of simulated signal and background events. The value of $m_t$ giving the best fit to *f* is then the estimated top mass. In Run I, CDF and DØ obtained:

$$m_t = 176.1 \pm 5.1 \text{ (stat.)} \pm 5.3 \text{ (sys.) GeV (CDF)}$$
$$173.3 \pm 5.6 \text{ (stat.)} \pm 5.5 \text{ (sys.) GeV (DØ)}$$

The largest systematic, for both experiments, is the uncertainty on the jet energy scale (~ 4 GeV).

CDF also measured the top mass in the 6-jets sample. Here there is a three-constraint fit, but the signal-to-background ratio is much worse. The result is

$$m_t = 186.0 \pm 10.0 \text{ (stat.)} \pm 5.7 \text{ (sys.) GeV (CDF)}$$

Again, the largest systematic is the uncertainty on the jet energy scale (4.4 GeV).

In the dilepton sample, things are harder, because two neutrinos are produced. Only four particles (two leptons, two jets) and two components of $E_T^{miss}$ are observed. There remain three constraints, so the system is underconstrained. A dynamical likelihood analysis is therefore performed. For each assumed value of the top mass, the kinematics can be reconstructed. An event weight is then calculated which parametrizes how probable it is that this event originated from a top-antitop of this given mass. The event weight distributions, as a function of $m_t$, are combined for all events, and compared with Monte Carlo samples as before. In Run I, CDF and DØ obtained:

$$m_t = 168.4 \pm 12.3 \text{ (stat.)} \pm 3.6 \text{ (sys.) GeV (DØ)}$$
$$167.4 \pm 10.3 \text{ (stat.)} \pm 4.8 \text{ (sys.) GeV (CDF)}$$

The largest systematic, for both experiments, is once again the uncertainty on the jet energy scale, but the systematics in this method are reduced compared to the lepton+jets channel.

The overall Tevatron average top mass is $m_t$ = 174.3 ± 5.1 GeV (CDF+DØ). In Run II, the data samples will be much larger. The systematic errors will need to be reduced to match, and this will require improved knowledge of the jet energy scale using data. This can come from Z+jet events, from photon+jet events, by using the W → jj decays within the top events, and by using Z → $\bar{b}b$ as a calibration sample. Gluon radiation effects will also need to be constrained using data, or reduced with harder cuts given a sufficient number of top events to afford the loss in statistics. Using double b-tagged events, again with a loss of statistics, we can greatly reduce the combinatorics. New mass extraction techniques are also being developed: DØ has reported a new, preliminary determination of the top mass using existing Run I lepton+jets data. The new technique makes use of more information per event, giving better discrimination between signal and background than the published 1998 analysis, and improves the statistical error equivalently to a factor 2.4 increase in the number of events.

With 2 fb$^{-1}$ of Run II data we will be able to drive the uncertainty on $m_t$ down to about 3 GeV/c$^2$ per experiment, and on $m_W$ down to about 25 MeV/c$^2$ per experiment. Measurements at these sensitivities will present a powerful test of the consistency of the standard model. With an ultimate capability of 15 MeV/c$^2$ for $m_W$ and 1–2 GeV/c$^2$ for $m_t$ per experiment, the indirect evidence for physics beyond the standard model would be compelling.

### 7.2. *Production Cross Section and Kinematics*

The top cross section is a test of QCD, and any discrepancy could possibly indicate new physics. In Run I, the measurement was performed in the lepton+jets, dilepton and all-jets final states. The results (5.9 ± 1.4 pb for DØ, 6.5 +1.7/−1.4 pb for CDF) are well within the range of QCD predictions (5−7 pb). The top quark transverse momentum is another test of the QCD prediction; in Run I, CDF found good agreement between data and expectations. In Run II we will also test for resonances in the top-antitop invariant mass that would signal the production of new particles decaying into top.

The Standard Model predicts that the top and antitop spins should be correlated. This is because their production is predominantly through quark-antiquark annihilation with a spin-1 s-channel gluon. Since the top lifetime ($4 \times 10^{-25}$ s) is less than the timescale of QCD hadronization ($1/\Lambda_{QCD} \sim$ few $\times 10^{-24}$ s), the top decays before the spin information is lost. The spin correlation can therefore be reconstructed from the decay products. The motivation to do this is to test the top quark's spin (is it really a spin-½ object as it should be?) and to check these assumptions about production and decay. The optimal spin-quantization basis is off-diagonal and has been derived by Mahlon and Parke [1]. Only like-spin combinations are produced in this reference frame. The analysis uses dilepton events, so in Run I it was very statistically limited; nevertheless, DØ carried the analysis through on the six available events. The result is consistent with expectations; with 2 fb$^{-1}$, it should be possible to distinguish between uncorrelated spins and the standard model expected correlation at better than the 2σ level.

The top is expected to decay predominantly (70%) into a longitudinally polarized W, with a 30% admixture of left-handed W's. The lepton $p_T$ distrribution can distinguish the various states. CDF find the Run I data are consistent with the standard model expectation, and the fraction of right-handed W's is consistent with zero.

### 7.3. *Single-top production*

Single top (top+bottom) production tests the electroweak properties of the top quark, and allows the Wtb coupling and the CKM element $|V_{tb}|^2$ to be extracted.

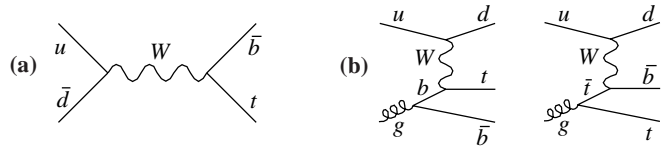

**Figure 9**: single top production

The cross section is about 0.7 pb for the s-channel process in Fig. 9 (a) and 1.7 pb for the t-channel process (b).

The signature consists of a W (lepton and missing $E_T$) together with two b-jets, one coming from top decay. In the s-channel process both b-jets are high-$p_T$ and central, while in the t-channel case the jet from top decay is still hard and central but the other is much softer, there also being a high-$p_T$ light quark jet (shown as a d-quark in Fig. 9(b)). It is desirable to separate the two processes based on these kinematic differences since they have different systematic errors and different sensitivities to new physics. The backgrounds for both are significantly higher than they are for top-antitop production (the W + 2 jet cross section is much higher than the W + 4 jet cross section).

In Run I, both experiments could only set limits (at several times the standard model level) on single top production. With 2 fb$^{-1}$ in Run II, we expect to see a clear signal. We will use it to measure the cross section to ~ 20%, to indirectly extract the width for t → Wb to ~ 25%, and thus $|V_{tb}|$ to ~ 12%. It is important to measure single top production since it is a background to many new physics searches; it is also sensitive to new dynamics like topcolor.

**8. The Higgs Boson**

In the standard model, the weak force is weak because the W and Z bosons interact with a scalar field (called the Higgs field) that permeates the universe with a finite vacuum expectation value of 246 GeV. If this same field couples to the fundamental fermions, it can explain their masses too. It should be possible to excite this field and observe its quanta—the long sought Higgs boson. It is the last piece of the standard model, and also a key to understanding any beyond-the-standard-model physics like supersymmetry. Finding it is a very high priority, but will require large datasets because the production cross section is low and the irreducible backgrounds are large. All of the properties of the Higgs are fixed in the standard model with the exception of its own mass: its couplings and decays are all determined.

There is no doubt that electroweak symmetry breaking occurs, so we know that something is coupling to the W and Z. We should remember, however, that the Higgs field need not result from a single, elementary scalar boson: there could be more than one particle (as occurs in SUSY), or composite particles could play this role (as in technicolor or topcolor). Whatever the field is, precision electroweak measurements tell us that it looks very much like a Standard Model Higgs (though its fermion couplings are less constrained). We also know that the WW cross section would violate unitarity at √s ~ 1 TeV without the existence of a Higgs; this is a real experiment accessible to the LHC, so there is no doubt that whatever plays the role of the Higgs will eventually be seen.

Over the last decade, the focus was on experiments at LEP. Direct searches excluded masses below 114 GeV, while indirect electroweak measurements combined with Tevatron top mass data in global fits to exclude Higgs masses above about 200 GeV. In mid 2000, there was excitement that hints of a signal for a Higgs with a mass of 115 GeV, right at the limit of sensitivity, had been seen at LEP. In the event, CERN management decided to shut off LEP operations in order to expedite construction of the LHC, and the resolution of the puzzle is now left to the Tevatron.

A relatively light Higgs, with a mass in the range between the current lower bound of 114 GeV and about 140 GeV, is produced with a cross section of about 1 pb at the Tevatron and decays mainly to $\bar{b}b$ quark pairs. This gives a signature of two b-jets, which is almost impossible to extract from the huge background of two b-jets from QCD processes. Instead, we will search for the production of a Higgs together with a W or Z boson. The cross section is a factor of 5 lower, but the W or Z decays a significant fraction of the time into an electron or a muon, and a high energy electron or muon is relatively easy to trigger on and isolate. We then have the simpler, but still very challenging, task of separating the signal of a vector boson plus a Higgs, from the backgrounds: vector bosons plus two b-jets, vector bosons plus a Z, with $Z \to \bar{b}b$, top pairs and single top. The $\bar{b}b$ mass resolution directly influences the signal significance; $Z \to \bar{b}b$ will serve as a calibration for $H \to \bar{b}b$.

As an example, consider $m_H$ = 115 GeV (interesting because of the LEP "hint"). In 15 fb$^{-1}$, we expect[2] 92 WH $\to$ lv $\bar{b}b$ events (with a background of 450); 90 ZH $\to$ vv $\bar{b}b$ events (with a background of 880); and 10 ZH $\to$ ll $\bar{b}b$ events (with a background of 44). These event rates would allow a 95% exclusion (if no signal is seen) with 2fb$^{-1}$ per experiment, and three and five standard deviation signals to be established with 5 and 15fb$^{-1}$ respectively. If we do see something, we will want to test whether it is really a Standard Model Higgs by measuring the production cross section (which is predicted for any given $m_H$), and searching for other decay modes. The modes $H \to \tau\tau$ and $H \to WW$ are expected to have branching ratios around 8–9% for $m_H$ = 115 GeV and might just be visible. $H \to \gamma\gamma$ should *not* be visible for a SM Higgs at the Tevatron. It will also be interesting to look for the process where the Higgs is produced in association with a top-antitop pair. The cross section is very low (a few fb) but the signal to background ratio is good, and one might see a signal at the few event level. This process tests the top quark Yukawa coupling.

At higher Higgs masses, above about 140 GeV, the dominant Higgs decay mode becomes W pairs. Here, the challenge is not to reconstruect b-jets, but to separate the H → WW signal from the other sources of W pairs. Angular cuts can be used to extract the Higgs signal from the WW background (the Higgs is spin-0 while a virtual Z is spin-1). With tight cuts one can obtain a signal to background of order 1 to 3.

All of these Higgs strategies and estimated sensitivities are taken from the Higgs and Supersymmetry workshop held at Fermilab in 1999 [2]. Right now, the experiments are developing the foundations needed to do this analysis for real in Run II: good jet resolution, high b-identification and trigger efficiencies, and a good understanding of all the backgrounds. These will also enable us to firm up these earlier estimates of Higgs sensitivity as a function of luminosity.

In the most popular extension to the Standard Model, supersymmetry (described below), there is still a Higgs boson: in fact the existence of a light Higgs is a very basic prediction of SUSY. These models contain an extended suite of Higgs particles: two neutral scalars h and H, a pseudoscalar A and a charged pair $H^{\pm}$. At tree level, the Higgs sector is decsribed by two parameters (usually taken to be $m_A$ and $\tan\beta$); on top of this, there are radiative corrections that depend on sparticle masses and the top mass. LEP has excluded $m_h$ < 91 GeV, $m_A$ < 92 GeV and $m_{H^{\pm}}$ < 79 GeV, and $\tan\beta$ < 2.4. Over much of the remaining allowed parameter space, the h is "light" (115−130 GeV) and the H, A and $H^{\pm}$ are much more massive. In this "decoupling limit", the h looks very much like the standard model Higgs and decays similarly, while the heavy Higgses tend to have enhanced decays to b's and taus. Tevatron searches for the standard Higgs therefore apply to its supersymmetric cousin h as well; if we exclude the existence of such a Higgs, we will have gone a long way towards excluding minimal supersymmetry at the electroweak scale. One area of Higgs physics that can be attacked with relatively modest luminosities (already feasible in 2003) is to search for the other supersymmetric Higgs bosons. Associated production of a neutral SUSY Higgs (h/H/A) together with a $b\bar{b}$ quark pair is enhanced for high but plausible values of $\tan\beta$, and tighter limits than those from LEP will already be reachable with a few hundred inverse picobarns. The Tevatron can also search for the production of charged Higgses in top decays: the decay to $H^+b$ competes with the standard model Wb decay, and would result in an enhanced production of tau leptons in top decays at large $\tan\beta$.

As long as we have adequate sensitivity, exclusion of a Higgs would still be a very important discovery for the Tevatron. In the Standard Model, we can only exclude the lower part of the mass range (up to 175 GeV with ~10 fb$^{-1}$), but this is the region favored by electroweak fits. In minimal SUSY, we can potentially exclude all of the allowed parameter space with ~5 fb$^{-1}$. There is also the possibility of discovering something other than a Higgs. In dynamical models like technicolor and topcolor, the role of the Higgs is played by a composite particle. Such models predict many other new particles in the mass range 100-1000 GeV, with strong couplings, large cross sections, and clear signatures at the Tevatron.

It is useful to remember what the Higgs search will and will not tell us. It will tell us what is the source of mass of the W and Z, and therefore why the weak force is weak. It will tell us the source of mass of the fundamental fermions. It will tell us whether there are fundamental scalars, and give a strong indication whether there is weak-scale SUSY (since a light Higgs is a basic feature of such models). It could tell us that there is other new physics at the weak scale, should we discover something like technicolor, or should we find a Higgs, but one whose mass, together with the top and W masses, is not consistent with precision electroweak fits. It will tell us the mass scale of new physics and point towards the machine we want to build after the linear collider. The search will not reveal why fermion masses have the values they do, nor what is the origin of flavor (what distinguishes a top quark from an up quark). It will not reveal what is the origin of mass in the universe, since the baryon masses are almost all due to QCD, while that of dark matter, if it turns out to be the lightest supersymmetric neutralino, is due to some unknown SUSY breaking mechanism. We don't even know what two thirds of the universes's mass—dark energy—is made of.

## 9. Searches for New Physics

As the world's highest energy collider, the Tevatron is the most likely place to directly discover a new particle or force. We know the standard model is incomplete; theoretically the most popular extension is to make it a part of a larger picture called supersymmetry (which is a basic prediction of superstring models). Here each known particle has a so-far unobserved and more-massive partner, to which it is related through a change of spin. If it exists, the lightest supersymmetric particle would be stable, and vast numbers of them would pervade the universe, explaining the astronomers' observations of dark matter. LEP has excluded sparticle masses below the 80–100 GeV range (and lightest neutralinos below 36 GeV); the Tevatron is now the only place to directly search for supersymmetry. There are a number of search strategies. The most copiously produced SUSY particles at a hadron collider would be the colored squarks and gluinos. These have cascade decays giving final states with missing energy ($E_T^{miss}$) plus jets (and often lepton(s) from charginos and neutralinos in the cascade). In Run I, squark and gluino masses below about 200 GeV were exluded (Fig. 10); with 2 fb$^{-1}$, this mass reach can be roughly doubled.

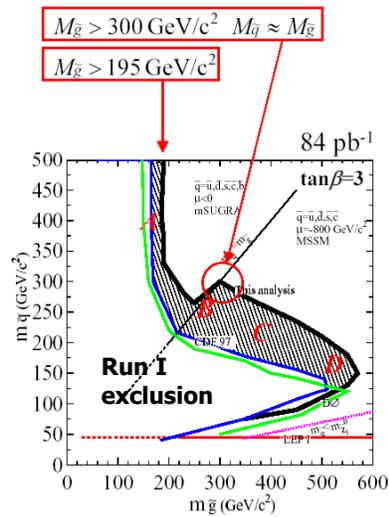

**Figure 10:** squark and gluino mass limits from Run I (CDF)

Another attractive search mode is for charginos and neutralinos through multilepton final states. Associated production of a chargino and a neutralino can give a trilepton signature for which the standard model background is very small. In Run I, this search was not competitive with LEP, but in Run II it will become increasingly important as squark and gluino production reaches its kinematic limits. The chargino mass reach is 150-180 GeV. We will also search for SUSY signatures with heavy flavor. Quite often the lightest squarks are the sbottom and stop, which can decay to b or c-jets. Gauge mediated SUSY models result in signatures with $E_T^{miss}$ + photon(s).

Searches for other new phenomena include leptoquarks, dijet resonances, new heavy W′ and Z′ bosons, massive stable particles, and monopoles.

The Tevatron allows us to experimentally test the new and exciting idea that gravity may propagate in more than four dimensions of spacetime. If there are extra dimensions that are open to gravity, but not to the other particles and forces of the standard model, then we could not perceive them in our everyday lives. But particle physics experiments at the TeV scale could see signatures such as a quark or gluon jet recoiling against a graviton, or indirect indications like an increase in high energy electron-pair production. These studies use the Tevatron to literally measure the shape and structure of space-time.

While it is good to be guided by theory, one should also remain open to the unexpected. Therefore both experiments have developed quasi-model-independent (signature-based) searches, which look for significant deviations from the Standard Model. In the Run I dateset, no significant evidence for new physics was found. Perhaps revealing different psychologies, DØ has quantified its agreement with the Standard Model at the 89% confidence level, while CDF has preferred to highlight some potential anomalies as worth pursuing early in Run II.

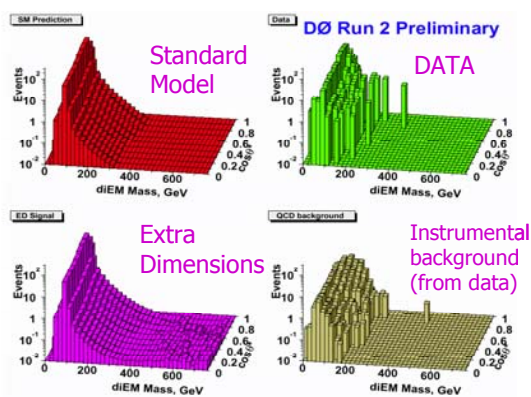

**Figure 11:** Run II search for virtual effects of extra dimensions in the pair production of photons and electrons (DØ)

The experiments have already embarked on a number of searches using Run II data. Work has started on understanding the $E_T^{miss}$ distribution in multijet events as a prelude to squark and gluino searches; trilepton candidates are also being accumulated. At DØ, a gauge-mediated SUSY search has set a limit on the cross section for $\bar{p}p \to E_T^{miss} + \gamma\gamma$. Also at DØ, virtual effects of extra dimensions are being sought in $e^+e^-$, $\mu^+\mu^-$ and $\gamma\gamma$ final states, and limits on the scale of new dimensions at the 0.9 TeV level can already be set. A search for leptoquarks decaying to electron+jet has been carried out. None of the cross sections or mass limits is better, yet, than published Run I results, but serves as a demonstration that the pieces are all in place.

**10. Physics Prospects for Run II**

At the same time as Fermilab's accelerator experts have been working to improve Tevatron operations, they have been trying to incorporate the lessons learned into a solid plan for the future. The planning for the accelerator complex is in two phases. The first focuses on US Fiscal Year 2003, which ends in September 2003. A full plan and schedule are in place. By the summer of 2003, each experiment should have recorded around 200 pb$^{-1}$ of Run II data (almost twice the Run I dataset). The centrepiece will be a greatly increased top quark sample, thanks to the higher beam energy and the much improved b-tagging capabilities of the detectors. A first look at $B_s$ mixing will be possible, together with lifetimes and branching ratio measurements from the B, $B_s$, $\Lambda_b$ and charm samples. Jet distributions at the highest energies will constrain proton structure, and searches will follow up on Run I anomalies and extend the Run I reach for many extensions to the standard model.

The second phase covers 2004 and beyond. It is now clear that it will take somewhat longer than had been anticipated to accumulate the large datasets ultimately foreseen for Run II: such is the price of realism. As long as the Tevatron remains the world's highest energy collider, it is a unique facility that must be exploited to the fullest; this will remain true until the LHC experiments start producing competitive physics results. If the LHC delivers first beam to ATLAS and CMS in 2008, we might expect a year to be spent commissioning the detectors and accelerator, with the first physics data by 2009. One year of concurrent running (2009-10) between LHC and Tevatron would not be unreasonable, since the low-mass Higgs searches at the two machines are complementary in several ways. With this in mind, we are preparing to run the Tevatron until the end of the decade in order to fully realize its physics potential.

The Run II physics program is a broad and deep one and will answer crucial questions about the universe. As Fig. 12 shows, there is no threshold at which this starts. There is compelling physics to be done each year and with each doubling of luminosity, starting now with a few hundred inverse picobarns and to the end of the decade with multi femtobarn data sets. To explore the 5 fb$^{-1}$ to 15 fb$^{-1}$ domain calls for upgrades to the CDF and DØ detectors. Primarily, these involve new trigger systems to handle more than ten interactions per crossing at the expected luminosity, and new silicon detectors that make use of LHC R&D to sustain the high radiation doses. These upgrades were successfully reviewed by the Department of Energy in September and are now moving towards approval with installation planned for 2005–6.

Run II is a marathon and not a sprint. The combination of high accelerator energy, excellent detectors, enthusiastic collaborations and data samples that are doubling every year guarantees interesting new physics results at each step. Each step answers important questions. Each leads on to the next. This is how we will lay the foundations for a successful LHC physics program—and hopefully a linear collider to follow.

**References**

1. G. Mahlon and S. Parke, Phys. Lett. **B 411**, 173 (1997).

2. M. Carena *et al*., Report of the Higgs Working Group of the Tevatron Run II SUSY/Higgs Workshop, hep-ph/0010338.

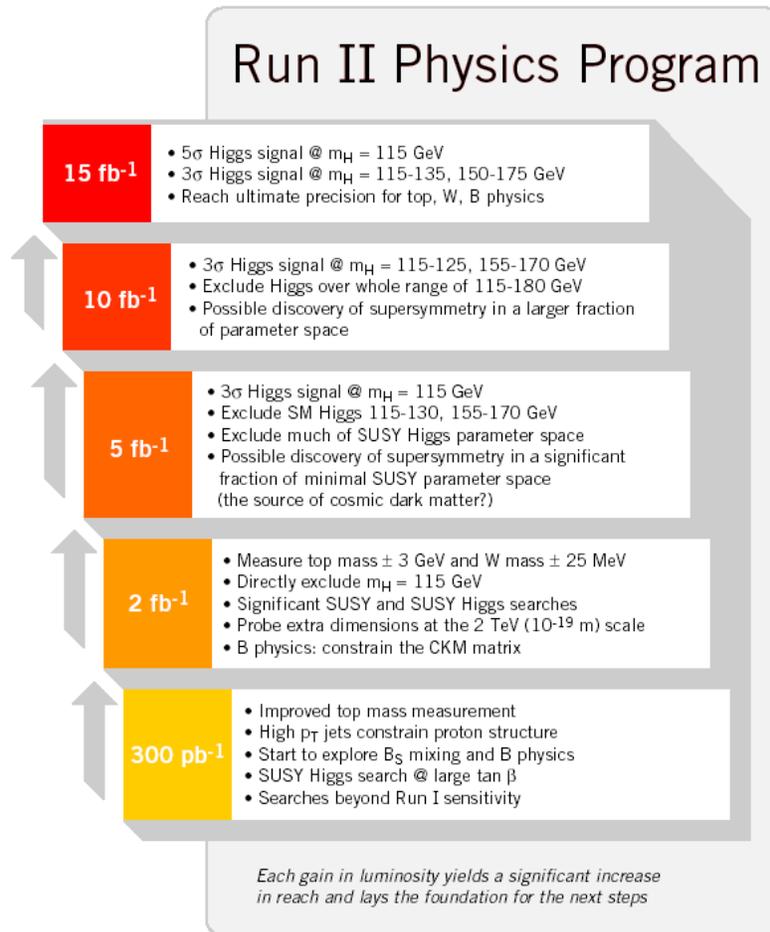

**Figure 12:** Summary of Run II physics reach for various integrated luminosities.

**Appendix: Student Exercise for TASI 2002**

**A Search for New Physics using DØ Run 1 Data**

*John Womersley, Fermilab*
*TASI Summer School, June 2002*

Goal of this exercise: explore the limits that can be placed on the production of technicolor particles using the data already taken at the Tevatron in 1992-95. Extrapolate to Run 2.

We will use the public interface to DØ Run 1 data, which is called Quaero.

**1. First, look at the PRL article describing Quaero:**

http://www-d0.fnal.gov/www_buffer/pub/pub_228.pdf or
http://www-d0.fnal.gov/www_buffer/pub/pub_228.ps

You'll see (in Table III of the paper) a number of channels where limits were set, including $\bar{p}p \to WH \to e\nu jj$. The paper did not, however, set limits on the technicolor process with a similar final state, $\bar{p}p \to W\pi_T \to e\nu jj$. It is suggested that this process may have a much higher cross section than the Higgs process, so it's interesting to see what can be done with the relatively low luminosity accumulated in Run 1.

This process, and a suggested search strategy, are described in:
http://arxiv.org/abs/hep-ph/9704455

**2. Go to the Quaero website**: http://quaero.fnal.gov

Look at the examples to see what one can do with this interface. You can ask Quaero to generate the signal events itself, using Pythia, or you can feed it an input file. In this case we'll ask it to use Pythia because Technicolor is implemented in Pythia.

The Pythia web page is http://www.thep.lu.se/~torbjorn/Pythia.html but a more useful short guide from LAPP can be found at http://wwwlapp.in2p3.fr/Pythia/. Page 28 and 29 of the short guide (PDF version) describe the technicolor implementation.

**3: Now let's set up our search.**

Under "signal:"
- Select the final state corresponding to evjj (e met 2j (nj))
- Select "smear" (we want the detector resolutions to be modelled)
- Select "Pythia input" since we want Quaero to generate the technicolor events itself.
- Enter the Pythia commands to generate $W\pi_T$ events: The easiest way is MSEL=50 which turns on all the technicolor processes.
- Click on "search" (that's what we want to do)
- Check the boxes corresponding to the backgrounds to be considered

Now under "variables:"
- If you like, enter a constraint (e.g. you could constrain the e and the missing $E_T$ to be a W). This is optional.
- Enter up to 3 variables in which Quaero will search for signal events: the hep-ph paper suggests the $p_T$ of the W, the $p_T$ of the dijet system, $\Delta\phi(jet_1, jet_2)$, and the mass of the dijet system.

Under "Requestor," enter your name and e-mail, and a short note describing what you are doing (please mention TASI).

**4. Now click "submit"!**

Stay logged in – if there are errors in the submission, Quaero will bounce an e-mail right back to you.

**5. Check your e-mail tomorrow** for the results.
See the examples on the following pages for typical output.

**6. What limit did you get?** How do this limit, and the Pythia cross section reported by Quaero, compare with the expected theoretical cross section from the hep-ph paper? Remember that Quaero quotes limits on (cross section × branching ratio), so you'll have to multiply by 9 to remove $B(W \to e\nu)$. How much more luminosity would it take to observe a signal in Run 2?

**7. If you have time, try optimizing** the constraints or variables to see if you can get a better limit on the same process. Try constraining the e and Missing $E_T$ to a W mass, for example, if you didn't already. Try other variables which might be better than mine.

Example of the plots that are produced:

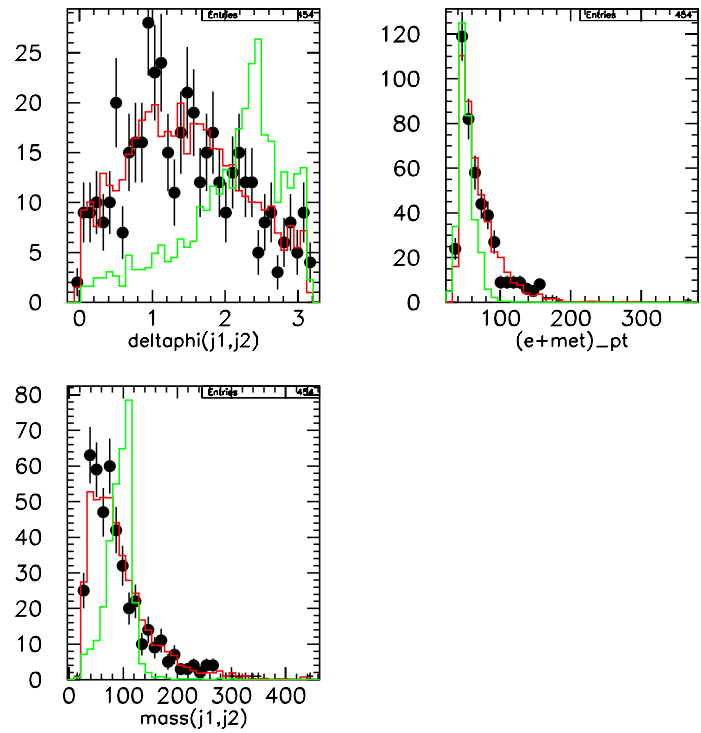

The plots show the three variables I specified in my search. The black points are the data, the red curve is the background, and the green is the signal expected multiplied by 50. You can see that the deltaphi and mass variables give quite good distinction between signal and backgound, but the $p_T$ variable isn't so great. Try to do better!

**Quaero FAQ**:
1. Yes, this is real collider data.
2. Yes, if you find a signal you can publish it.